\documentclass[article]{IEEEtran}
\usepackage[utf8]{inputenc}
\usepackage{xcolor}
\usepackage{booktabs}
\usepackage{blindtext, graphicx}
\usepackage{amsmath}
\usepackage{textcomp}
\usepackage{graphicx} 
\usepackage{float}
\usepackage{caption}
\usepackage{amsmath,amssymb}  
\usepackage{paralist}
\usepackage{titlesec}  
\usepackage{parskip}   
\usepackage{indentfirst}
\usepackage{url}

\usepackage{breakurl}

\titlespacing*{\section}{0pt}{3.5ex plus 1ex minus .2ex}{2.3ex plus .2ex}
\titlespacing*{\subsection}{0pt}{3.25ex plus 1ex minus .2ex}{1.5ex plus .2ex}

\usepackage[shortlabels]{enumitem}
\usepackage{bm}  
\usepackage{multirow}
\usepackage[caption=false,font=footnotesize]{subfig}
\usepackage[pdftex,bookmarks,colorlinks=false,breaklinks,hidelinks]{hyperref}  
\usepackage{memhfixc} 
\usepackage[mathscr]{euscript}
\DeclareSymbolFont{rsfs}{U}{rsfs}{m}{n}
\DeclareSymbolFontAlphabet{\mathscrsfs}{rsfs}
\title{Risk Assessment of Distribution Networks Considering Climate Change and Vegetation Management Impacts
 \thanks{D.~ Zhao, U.~T.~Salman and Z.~Wang are with the Department of Electrical and Computer Engineering at the University of Connecticut, Storrs, CT, USA 06268.  ($^{\dagger}$Corresponding author: Dr. Zongjie Wang, zongjie.wang@uconn.edu). }%
 \thanks{This work was funded in part by the Department of Energy funded project entitled `` Wind Impact Study for Power Resilience''}%
 }
 
\author{Di Zhao, \textit{Student Member},  Umar Salman, \textit{Student Member}, and Zongjie Wang$^\dagger$\vspace{-24pt}, \textit{Senior Member}}
\usepackage{cite}
\usepackage{graphicx}
\usepackage{nomencl}
\makenomenclature

\providetoggle{nomsort}
\settoggle{nomsort}{true} 

\makeatletter
\iftoggle{nomsort}{%
    \let\old@@@nomenclature=\@@@nomenclature        
        \newcounter{@nomcount} \setcounter{@nomcount}{0}%
        \renewcommand\the@nomcount{\two@digits{\value{@nomcount}}}
        \def\@@@nomenclature[#1]#2#3{
          \addtocounter{@nomcount}{1}%
        \def\@tempa{#2}\def\@tempb{#3}%
          \protected@write\@nomenclaturefile{}%
          {\string\nomenclatureentry{\the@nomcount\nom@verb\@tempa @[{\nom@verb\@tempa}]%
          \begingroup\nom@verb\@tempb\protect\nomeqref{\theequation}%
          |nompageref}{\thepage}}%
          \endgroup
          \@esphack}%
      }{}
\makeatother

\begin{document}

\maketitle
\begin{abstract}
This paper presents a comprehensive risk assessment model for power distribution networks with a focus on the influence of climate conditions and vegetation management on outage risks. Using a dataset comprising outage records, meteorological indicators, and vegetation metrics, this paper develops a logistic regression model that outperformed several alternatives, effectively identifying risk factors in highly imbalanced data. Key features impacting outages include wind speed, vegetation density, quantified as the enhanced vegetation index (EVI), and snow type, with wet snow and autumn conditions exhibiting the strongest positive effects. The analysis also shows complex interactions, such as the combined effect of wind speed and EVI, suggesting that vegetation density can moderate the impact of high winds on outages. Simulation case studies, based on a test dataset of 618 samples, demonstrated that the model achieved an 80\% match rate with real-world data within an error tolerance of \(\pm 0.05\), showcasing the effectiveness and robustness of the proposed model while highlighting its potential to inform preventive strategies for mitigating outage risks in power distribution networks under high-risk environmental conditions. Future work will integrate vegetation height data from Lidar and explore alternative modeling approaches to capture potential non-linear relationships.

\end{abstract}
\vspace{-10pt}  

\section{Introduction}
\label{sec:introduction}
\setlength{\parindent}{2em}  

\indent Extreme weather has a substantial impact on the economy, infrastructure, and individuals~\cite{donatti2024}. Overhead distribution networks are particularly vulnerable to natural disasters such as windstorms, snowstorms, or hurricanes \cite{younesi2022investigating,younesi2023pathway}. This vulnerability is often exacerbated by the secondary impacts such as fallen trees or branches \cite{salman2024power}. In fact, in many overhead distribution networks, tree limbs and trees are among the leading causes of power outages. Specifically, trees account for 90\% of power outages caused by extreme weather events~\cite{eversource}. Other factors include lightning, equipment failures, and degradation, many of which are triggered or worsened by extreme event conditions. Among these, extreme weather events such as high wind speeds are often the primary drivers. A resilience integrated frameworks that incorporates tree failure model in extreme events is investigated in \cite{salman2024power}.

Vegetation management serves as the primary strategy for mitigating the risk of outages caused by tree-related impacts on overhead distribution networks~\cite{vm}. By establishing a clear understanding of the relationship between extreme weather conditions, vegetation characteristics, and outage occurrences, utilities can identify high-risk areas more accurately. This enables the implementation of targeted or prioritized vegetation management strategies, such as trimming or removing trees in critical zones. Such an approach reduces outage rates while minimizing maintenance costs and enhancing operational efficiency. Furthermore, incorporating advanced data analytics, such as vegetation indices or LiDAR-based assessments, can refine these efforts by identifying specific vegetation attributes, such as density, height, and proximity to power lines, that contribute to outage risks. Ultimately, this data-driven vegetation management framework supports proactive decision-making and ensures more resilient power distribution systems in the face of extreme weather conditions.

Previous studies have established various machine learning models to predict power outages. For example, time series and regression models have been developed to consider the impact of vegetation growth and climate change on outages~\cite{milan}. 
Additionally, predictive capabilities have been enhanced by integrating LiDAR data, which provides detailed vegetation height and density information in paper~\cite{wanik}. While these studies have advanced the field, they often focus on specific aspects, such as storm risks or vegetation metrics, without fully addressing the interactions between meteorological and vegetation factors and many models also struggle with imbalanced outage data. This highlights the need for an integrated approach to analyze combined weather and vegetation effects while ensuring robustness in imbalanced scenarios.

\begin{figure*}[htbp]
    \centering
    \includegraphics[width=.8\linewidth]{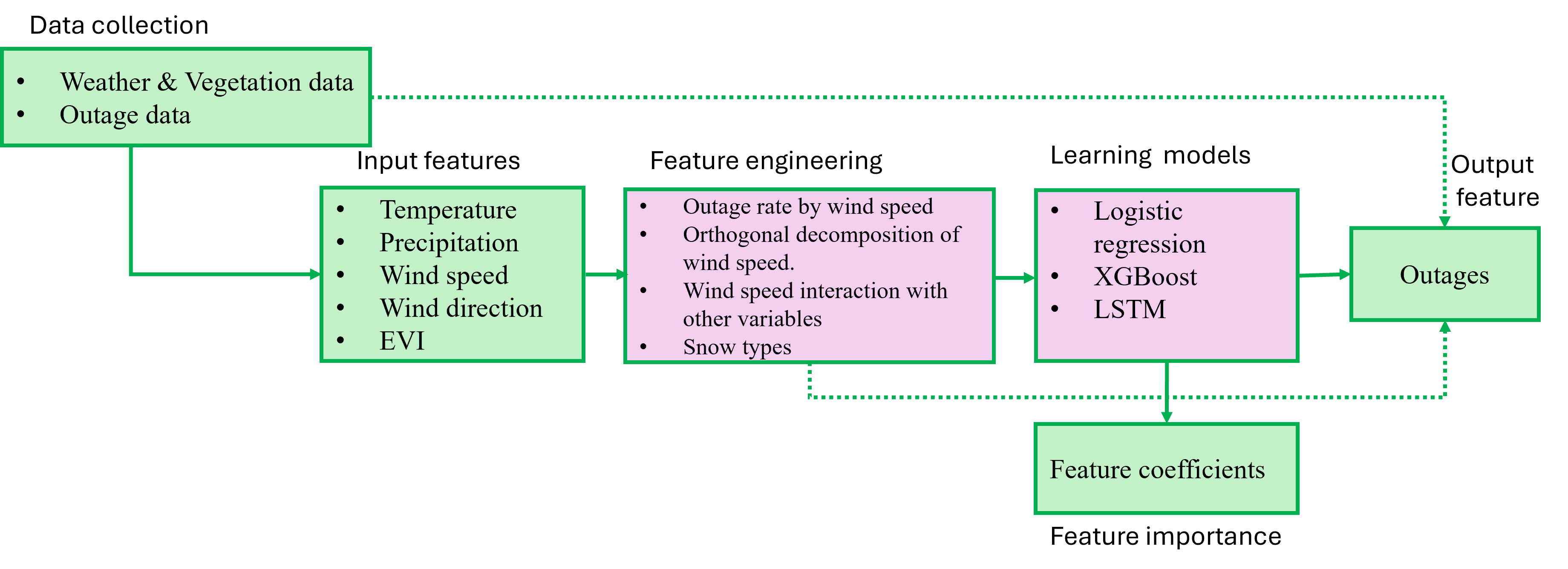} 
    \vspace{-15pt}
    \caption{Simulation framework.}
    \label{fig:framework}
\end{figure*}

Motivated by these observations, the main contributions of this paper are shown as follows:
\begin{itemize}
    \item Develop a robust risk assessment model that integrates meteorological and vegetation factors to predict power outages and address critical gaps in existing studies.
    \item Capture complex interactions between key variables, such as wind speed and vegetation density by using advanced metrics such as the Enhanced Vegetation Index (EVI).
    \item Address the challenges of imbalanced outage data by implementing a logistic regression framework optimized for identifying significant risk factors to ensure both accuracy and interpretability.
    \item Validate the model through simulation case studies on a test dataset of 618 samples, capturing critical trends across varying environmental conditions, such as high wind speeds, dense vegetation, and wet snow, using real-world data to demonstrate its effectiveness and practical applicability in identifying high-risk areas for outages.
\end{itemize}



\section{Methodology}
\label{sec:method}
This analysis focuses on vegetation-related outages caused by branch contact or tree falls during extreme weather. Key factors include meteorological indicators (e.g., temperature, precipitation, wind speed, and wind direction) and vegetation metrics. The interactions among these variables are also analyzed for their combined impact on the risk of outages.

Fig.~\ref{fig:framework} 
outlines the simulation framework, including data collection, feature engineering, model selection, and feature importance extraction, detailed in the following subsections.

\subsection{Data Collection}
\subsubsection{Outage Events}
The outage records, provided by National Grid, cover a specific power line over the past 10 years. The dataset includes details such as date, cause, location, duration, and affected customers. For this study, only vegetation-related outages, such as branch contact and tree falls, were included. Over the past 10 years, there have been about 149 tree-related outages.

\subsubsection{Meteorological Features}
Weather data was collected from a Meteostat station located near the power line, providing daily records of variables such as average temperature, precipitation, wind speed, and wind direction. The dataset contains a minimal amount of missing values (less than 1\%), which were imputed using the average of the previous three days' records to ensure data continuity and integrity.
\subsubsection{Vegetation Feature}
The enhanced vegetation index (EVI), sourced from the Moderate Resolution Imaging Spectroradiometer (MODIS) and the Visible Infrared Imaging Radiometer Suite (VIIRS), measures vegetation health and density, with values typically ranging from -1 to 1. Higher EVI values indicate denser, healthier vegetation. To align with daily weather and outage data, EVI values recorded every 16 days were linearly interpolated to a daily resolution. Missing data after mid-May 2024 were filled using the average EVI for the same dates from 2016–2023, preserving seasonal trends for robust analysis.

\begin{figure}[h]
    \centering
    \includegraphics[width=0.8\linewidth,height=4.0cm]{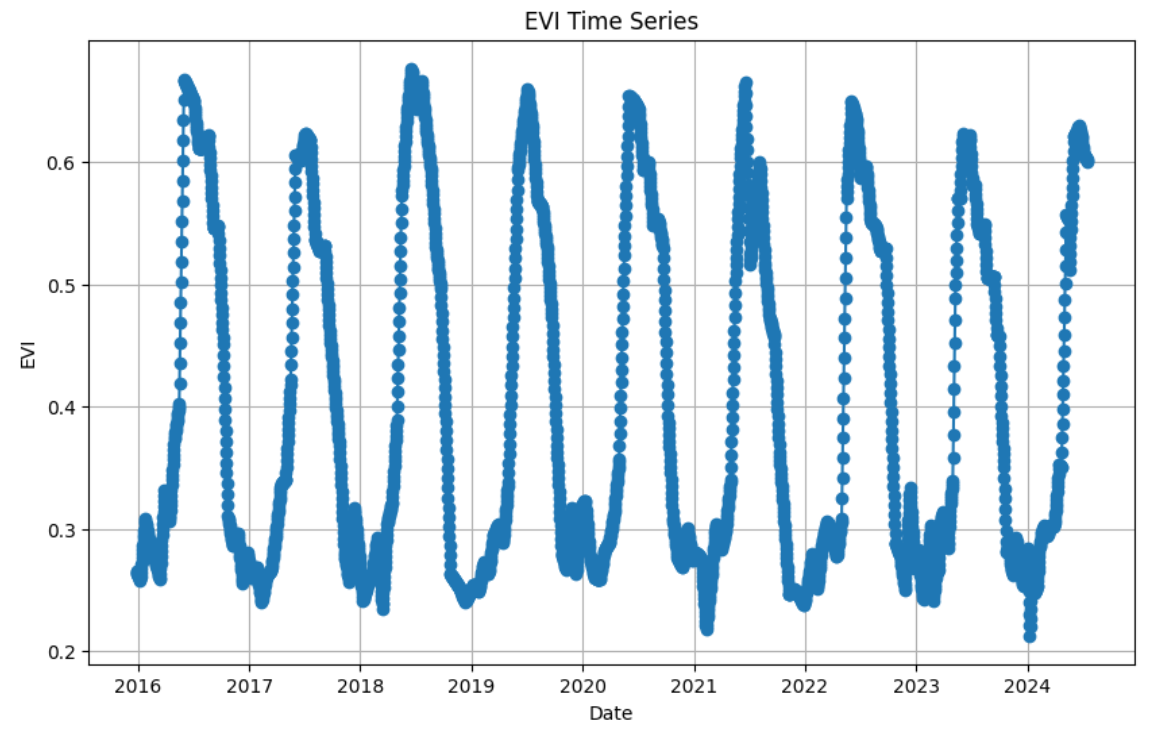} 
    \caption{EVI trends from Jan 2016 through Jul 2024.}
    \label{fig:2016-2024 EVI}
\end{figure}

 Fig.~\ref{fig:2016-2024 EVI} exhibits a seasonal pattern of EVI, with higher values observed during spring and summer, indicating dense vegetation coverage and healthy foliage. As the seasons transition into fall and winter, the EVI values decline, reaching a low point that reflects leaf drop and a more fragile state of vegetation.

\subsection{Feature Engineering}
\subsubsection{Outage Rate by Wind Speed Bins}
When analyzing the relationship between wind speed and outages, we examined the wind speed distribution and outage rate.

\begin{figure}[h]
    \centering
    \includegraphics[width=1\linewidth,height=4cm]{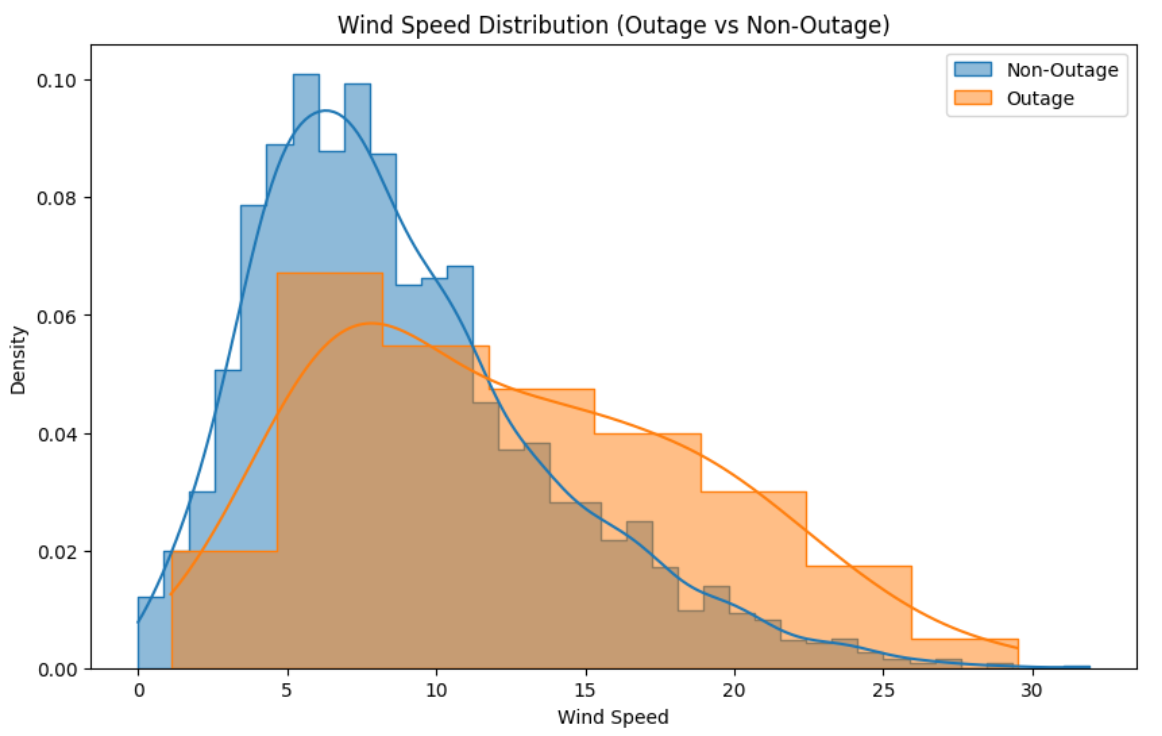} 
    \caption{Wind speed distributions illustrating the comparison between outage and non-outage events.
}
    \label{fig:wind_speed}
\end{figure}

Fig.~\ref{fig:wind_speed} shows the wind speed distributions for outage and non-outage events. The non-outage distribution peaks around 5 m/s, while the outage distribution is more spread across higher wind speeds, with peaks observed from 5 m/s to 30 m/s. This highlights the increased vulnerability of the power network under high wind conditions.

Additionally, we categorized the wind speed data into the following bins: 0–5 m/s, 5–10 m/s, 10–15 m/s, 15–20 m/s, 20–25 m/s, and 25+ m/s. This binning approach allows for a detailed examination of how outage likelihood varies across different wind speed ranges. 

\begin{figure}[h]
    \centering
    \includegraphics[width=0.8\linewidth,height=4.5cm]{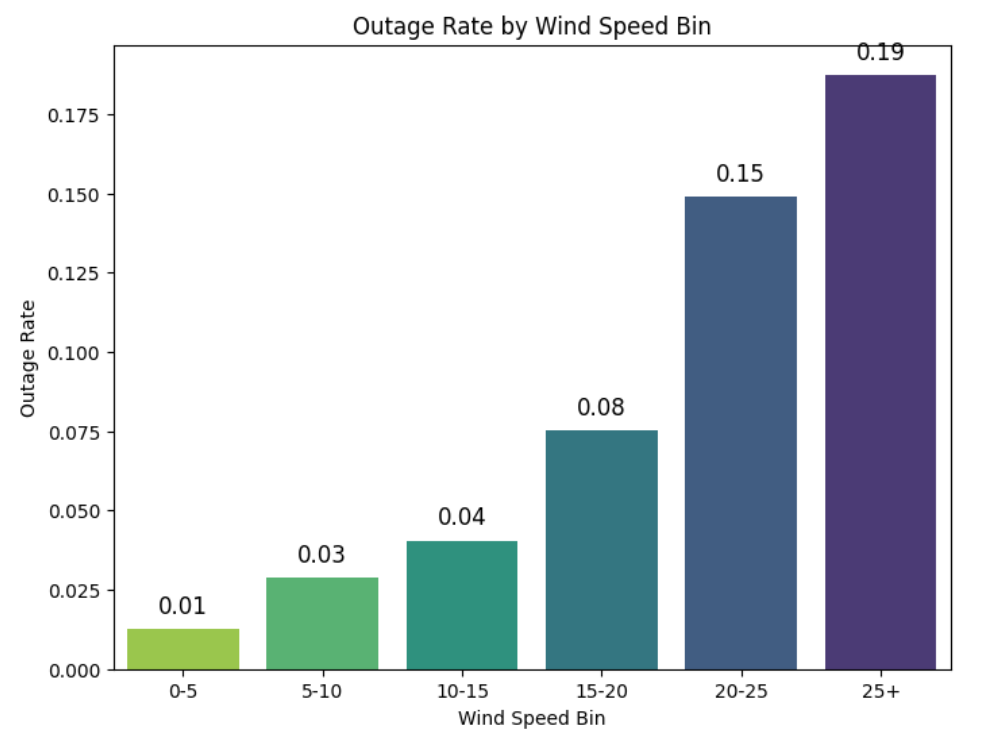} 
    \caption{Outage rates by wind speed bins.}
    \label{fig:outage_rate_by_windspeed}
\end{figure}

In Fig.~\ref{fig:outage_rate_by_windspeed}, The X-axis represents wind speeds(B) and the Y-axis depicts the outage rate, calculated as shown in Equation~\ref{eq:outage_rate}. The visualization reveals a strong positive correlation between wind speed and outage rates, with higher wind speeds leading to progressively increased outage rates. This trend underscores the heightened vulnerability of the power distribution network under extreme wind conditions and highlights the importance of incorporating wind speed as a key feature in predictive modeling. 

\begin{equation}
\text{Outage Rate}_{\text{bin}} = P(\text{Outage} | B) = \frac{N_{\text{out}, B}}{N_{\text{total}, B}} \label{eq:outage_rate}
\end{equation}
Where:
\begin{itemize}
    \item $B$: Wind speed bin (e.g., 0-5 m/s, 5-10 m/s, ...).
    \item $N_{\text{out}, B}$: Number of outage days within the wind speed bin $B$.
    \item $N_{\text{total}, B}$: Total number of days within the wind speed bin $B$.
    \item $P(\text{Outage} | B)$: Outage rate within wind speed bin $B$).
\end{itemize}

\subsubsection{Snow Type}
In the original weather data, there was no explicit information regarding snow conditions. To address this, this paper derives snow-related conditions using precipitation and temperature data, and categorizes them into the following types: 
\begin{itemize}
    \item \textbf{No Snow}: precipitation is zero \\
    \item \textbf{Dry Snow}: Precipitation \(>0\)  and temperature \(<0\)°C, representing less dense, moderate-risk snow~\cite{snowtype}\\
    \item \textbf{Wet Snow}: Precipitation \(>0\) and temperature between 0°C and 2°C, characterized by high density and adhesion~\cite{snowtype}
\end{itemize}


\begin{figure}[h]
    \centering
    \includegraphics[width=0.8\linewidth,height=4.0cm]{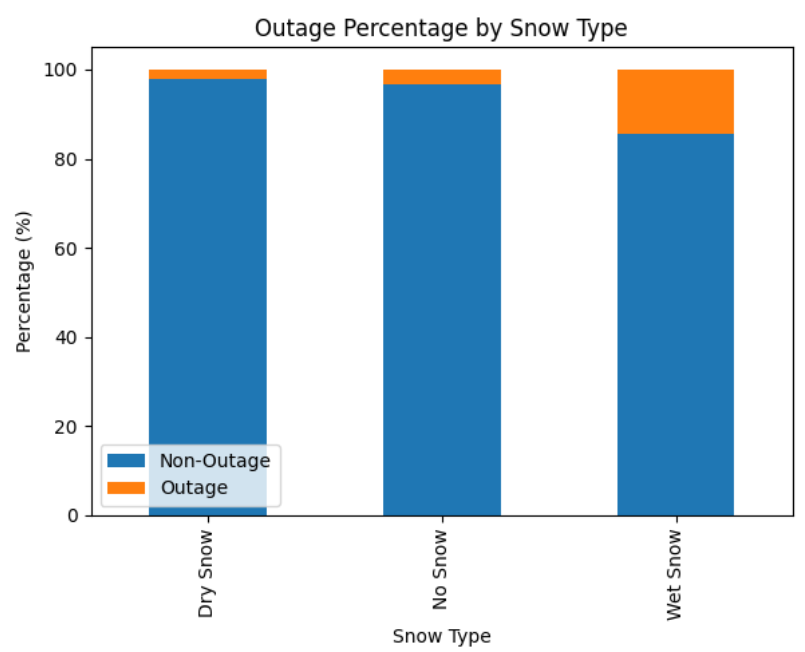} 
    \caption{Outage rates by snow types.}
    \label{fig:outage_rate_by_type}
\end{figure}
\vspace{-10pt}
Fig.~\ref{fig:outage_rate_by_type} illustrates that the probability of outages under wet snow conditions is significantly higher compared to the other two conditions (no snow and dry snow). This result shows the heightened vulnerability of the power system in wet snow environments, where the increased weight and adhesion of snow on branches and power lines lead to a greater risk of structural damage and subsequent outages.

\subsubsection{Orthogonal decomposition of wind speed}
The original wind direction data (0–360°) represents the angle of wind speed. To enhance analysis, wind speed was decomposed into two orthogonal components: east-west and north-south. This accounts for the varying impact of wind on power lines and vegetation based on orientation. Analyzing these components improves the representation of wind's directional effects on outage risk, enhancing predictive modeling.
\begin{itemize}
    \item \textbf{Horizontal}: \(V_x = |\vec{W}| \cdot \cos{\theta}\)
    \item \textbf{Vertical}:  \(V_y = |\vec{W}| \cdot \sin{\theta}\)
\end{itemize}
Where $\vec{W}$ denotes the direction of wind speed.
\subsubsection{Interaction of Variables}

\begin{itemize}
    \item \textbf{wind speed * \( V_x \) / \( V_y \)}: Represents the interaction of wind speed with its horizontal (\( V_x \)) and vertical (\( V_y \)) components.
    \item \textbf{wind speed * temperature}: 
    Examines the combined effect of wind and temperature on outage risk, reflecting potential system stress under extreme conditions.

    \item \textbf{wind speed * EVI}: 
    Explores the potential effects of high wind speeds in densely vegetated areas, where the interaction of strong winds with dense vegetation may 
    influence power outages.
\end{itemize}

\subsection{Modeling Approach}

In this analysis, we experimented with logistic regression, XGBoost, and Long Short-Term Memory (LSTM) models. Equation~\ref{eq:logistic_regression} is the mathematics expression of the logistic regression which is the best performance model in this analysis~\cite{lr}. During the data pre-processing phase, we extracted seasonal information from the date variable, applied one-hot encoding to categorical variables, and used standard scaling on all continuous variables. Standard scaling, as shown in Equation~\ref{eq:scaling}, transforms continuous variables by subtracting the mean (\(\mu\)) and dividing by the standard deviation (\(\sigma\)), standardizing the values to fall within a range centered around 0 with a standard deviation of 1. This transformation ensures that the importance of variables is not affected by differences in units, which is critical because many machine learning algorithms are sensitive to the scale of input features~\cite{scaling}. Without standard scaling, features with larger numerical ranges could dominate the model’s learning process, leading to suboptimal performance.
\vspace{-10pt}
\begin{equation}
\log\left(\frac{P}{1 - P}\right) = \alpha + \sum_{i=1}^{n} \beta_i X_i
\label{eq:logistic_regression}
\end{equation}
\vspace{-10pt}
\begin{equation}
X_{\text{scaled}} = \frac{X - \mu}{\sigma}
\label{eq:scaling}
\end{equation}

Considering the unbalance of data, the resampling technique was applied to correct this problem, otherwise the model is more inclined to learn and predict the majority group, resulting in poor prediction ability of the model for minority group. SMOTEENN~\cite{chawla2002smote}~\cite{wilson1972enn} was used in this analysis, which is a combination of SMOTE~\cite{chawla2002smote} and Edited Nearest Neighbors~\cite{wilson1972enn}. This approach integrates both synthetic sampling for the  minority class and noise reduction in majority class, so that the model can learn each class balanced.

Given the temporal nature of meteorological data, the dataset was split by selecting the first 80\% for the training set and the remaining 20\% for the testing set. This split helps prevent data leakage, which could otherwise lead to overly optimistic performance estimates by allowing future information to influence the training process.

The LSTM model consists of two LSTM layers with dropout to prevent overfitting, followed by a dense layer with a sigmoid activation function for binary classification. It was trained with 20 epochs and a batch size of 64. The XGBoost model included hyperparameters like max depth, learning rate, number of estimators, and scale positive weight, with LASSO regularization added to enhance generalization. The model was optimized using grid search with F1 score.

\section{Results}
\vspace{-10pt}
\label{sec:results}
\subsection{Test System and Model performance}

The analysis focuses on a Massachusetts distribution network with 78 km of overhead and 1 km of underground lines. Serving a rural, forested area with dense vegetation, the network is highly vulnerable to tree-related outages from high winds and heavy snowfall.


First, we evaluate the predictive accuracy and robustness of the regression models. Next, predicted results from 618 real-world records are compared against actual data. The coefficients of each feature are analyzed to assess their significance and impact. Finally, we discuss the implications of these findings for the power system, highlighting key vulnerabilities and strategies to mitigate outage risks.

Table~\ref{tab:model_performance} compares the performance of Logistic Regression, XGBoost, and Long Short-Term Memory (LSTM) models across key metrics. Precision reflects the accuracy of positive predictions, while recall indicates how well the model identifies actual positive cases. F1 score balances precision and recall, and AUC evaluates the model's ability to distinguish between positive and negative cases across thresholds. Logistic Regression achieves the best overall performance with an F1-Score of 0.87 and a ROC-AUC of 0.73, demonstrating its robustness in capturing minority-class predictions in the imbalanced dataset. This highlights its suitability for the outage prediction task.

\begin{figure}[h]
    \centering
    \includegraphics[width=.8\linewidth]{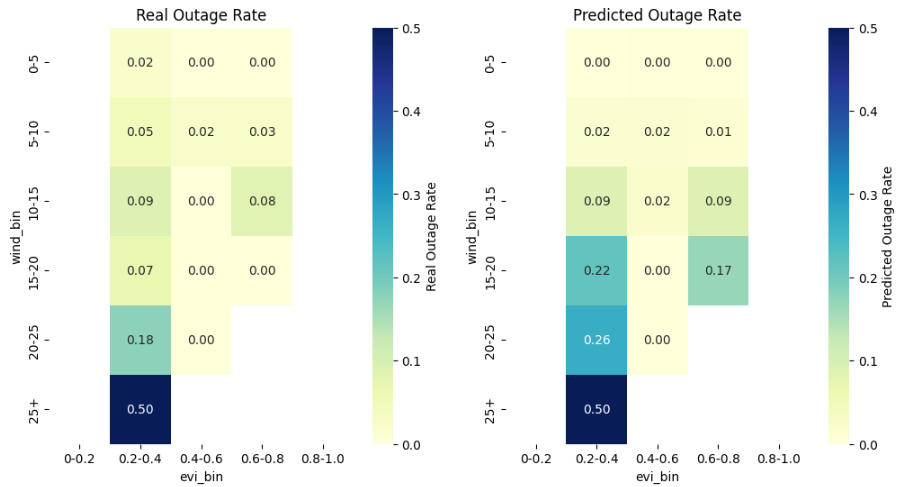} 
    \caption{Heatmap of Predicted vs. Actual Outage Rates.}
    \label{fig:heatmap}
\end{figure}

Fig~\ref{fig:heatmap} illustrates the outage rates across wind speed and vegetation bins. The left panel shows the observed outage rates, while the right panel depicts the predicted rates. The model captures key patterns in the data, particularly in bins with high wind speeds and dense vegetation, demonstrating strong agreement with real-world trends.  
\vspace{-15pt}

\begin{table}[h]
\centering
\caption{Model Performance Comparison}
\label{tab:model_performance}
\begin{tabular}{lcccc}
\toprule
\textbf{Model} & \textbf{Precision} & \textbf{Recall} & \textbf{F1-Score} & \textbf{ROC} \\
\midrule
Logistic Regression & 0.94 & 0.83 & 0.87 & 0.73 \\
XGBoost             & 0.93 & 0.72 & 0.80 & 0.69 \\
LSTM                & 0.95 & 0.80 & 0.86 & 0.70 \\
\bottomrule
\end{tabular}
\end{table}
\vspace{-15pt}

\vspace{-15pt}
Table II presents the logistic regression model coefficients, ranked by absolute value to indicate feature importance in predicting outages. Key factors include season, snow type, wind speed (wspd), and EVI, which significantly impact the model's predictions. Seasonal patterns highlight increased outage risk during autumn and winter, while wet snow, with the highest positive coefficient, underscores its severe impact on power infrastructure. Wind speed and EVI further demonstrate the critical roles of weather and vegetation density in outage risk. These results have validated the model’s ability to capture key risk drivers and provide actionable insights for targeted mitigation strategies.

\begin{table}[H]
\centering
\caption{Feature importance.}
\label{tab:feature_importance}
\begin{tabular}{|l|c|c|}
\hline
\textbf{Feature} & \textbf{Coefficient} & \textbf{Abs\_Coefficient} \\ \hline
snow\_type\_Wet\_Snow & 3.139 & 3.139 \\
season\_autumn & 2.952 & 2.952 \\
season\_summer & 2.077 & 2.077 \\
season\_winter & 1.811 & 1.811 \\
wspd & 1.500 & 1.500 \\
snow\_type\_No\_Snow & 1.117 & 1.117 \\
EVI & 0.857 & 0.857 \\
ws\_evi & -0.535 & 0.535 \\
prcp & 0.532 & 0.532 \\
wind\_x & -0.428 & 0.428 \\
cos\_wdir & -0.226 & 0.226 \\
wind\_y & -0.204 & 0.204 \\
tavg & 0.160 & 0.160 \\
wind\_temp & 0.156 & 0.156 \\
sin\_wdir & 0.078 & 0.078 \\ \hline
\textbf{Intercept} ($\alpha$) & -4.073 & - \\ \hline
\end{tabular}
\end{table}
\vspace{-15pt}
\subsection{Discussion}
\label{subsec:discussion}
In this analysis, logistic regression emerged as the best-performing model and has demonstrated a strong capability to distinguish between classes, even with the minority class accounting for less than 4\% of the data. The coefficients for wet snow and autumn are the highest, signifying that these conditions substantially increase the probability of outages compared to others. Following these, wspd and the EVI have also shown relatively high coefficients, indicating that higher wind speeds or denser vegetation significantly elevate outage risks. Additionally, rainfall exhibits a positive relationship with outages, further highlighting the role of adverse weather conditions. The interaction term between wspd and EVI reveals a negative coefficient. When EVI values are low to moderate, wind speed exerts a stronger influence on outage risk. However, as EVI increases, denser vegetation appears to moderate the effect of wind speed, leading to a reduced combined impact. This simulation result suggests that the relationship between wind speed and vegetation density is not purely additive but instead reflects complex environmental dynamics.
Coefficients of wind direction components (\textit{cos\_wdir}, \textit{sin\_wdir}, \textit{wind\_x}, \textit{wind\_y}) show that wind direction has a minor influence on outage risk. While certain directions may slightly reduce the likelihood of outages, their impact is secondary to dominant factors like wind speed and vegetation, likely depending on local environmental conditions.

Overall, the logistic regression model captures these intricate relationships effectively and offers valuable insights into the interplay between weather, vegetation, and outage risk. These results have highlighted the importance of prioritizing key factors such as wet snow, wind speed, and vegetation density in mitigation strategies while considering effects of interactions and directional components.
\vspace{-10pt}
\section{Conclusion}
\label{sec:conclusion}
This paper assesses the risk of distribution network outages by analyzing the impact of climate conditions and vegetation management. Using features engineered from historical outage data, meteorological indicators, and vegetation metrics, a logistic regression model was developed and compared against other models for both balanced and imbalanced datasets. The analysis identified wind speed, enhanced vegetation index (EVI), and snowstorms as key factors influencing outages. Additionally, interactions between wind speed and EVI revealed that vegetation density can moderate the impact of high winds on outages; however, their combined effect is not purely additive. The simulation results have provided valuable insights for managing outage risks under high-risk environmental conditions. Future work could integrate vegetation height and density from LiDAR data to enhance the model and explore alternative approaches to capture non-linear relationships. The results in this paper have practical implications, which enable better preparation for high-risk conditions, particularly during weather events. The proposed model can be integrated into operational decision-making by supporting preventive measures to reduce disruptions and enhance the resilience of power distribution networks.

\vspace{-15pt}

\end{document}